\documentclass[aps,preprint,preprintnumbers,amsmath,amssymb,amsfonts,groupedaddress,superscriptaddress,showpacs,showkeys,floatfix]{revtex4-1}

\usepackage{graphicx}
\usepackage{subfigure}
\usepackage{bm}
\usepackage{amsmath}
\usepackage{color}
\usepackage{array}
\usepackage{CJK}

\newcommand{\bmath}{\begin{mathletters}}
\newcommand{\emath}{\end{mathletters}}
\newcommand{\be}{\begin{eqnarray}}
\newcommand{\ee}{\end{eqnarray}}
\newcommand{\ba}{\begin{array}}
\newcommand{\ea}{\end{array}}

\newcommand{\no}{\nonumber}

\newcommand{\bt}{\beta}

\newcommand{\h}{\hbar}
\newcommand{\pr}{\prime}

\newcommand{\calI} {\mathcal I}

\newcommand{\calP} {\mathcal P}

\newcommand{\STA} {\mathrm{STA}}
\newcommand{\ad}  {\mathrm{ad}}
\newcommand{\cd}  {\mathrm{cd}}

\begin{document}
\title{Experimental demonstration of work fluctuations along a shortcut to adiabaticity with a superconducting Xmon qubit}

\author{Zhenxing Zhang}
 \affiliation{Physics Department, Zhejiang University, Hangzhou, 310027, China}
\author{Tenghui Wang}
 \affiliation{Physics Department, Zhejiang University, Hangzhou, 310027, China}
 \author{Liang Xiang}
 \affiliation{Physics Department, Zhejiang University, Hangzhou, 310027, China}
 \author{Zhilong Jia}
 \affiliation{Key Laboratory of Quantum Information, University of Science and Technology of China, Hefei, 230026, China}
 \author{Peng Duan}
 \affiliation{Key Laboratory of Quantum Information, University of Science and Technology of China, Hefei, 230026, China}
 \author{Weizhou Cai}
 \affiliation{Center for Quantum Information, Institute for Interdisciplinary Information Sciences, Tsinghua University, Beijing, 100084, China}
 \author{Ze Zhan}
 \affiliation{Physics Department, Zhejiang University, Hangzhou, 310027, China}
 \author{Zhiwen Zong}
 \affiliation{Physics Department, Zhejiang University, Hangzhou, 310027, China}
 \author{Jianlan Wu }
 \email{jianlanwu@zju.edu.cn}
 \affiliation{Physics Department, Zhejiang University, Hangzhou, 310027, China}
 \author{Luyan Sun}
 \affiliation{Center for Quantum Information, Institute for Interdisciplinary Information Sciences, Tsinghua University, Beijing, 100084, China}
 \author{Yi Yin}
 \email{yiyin@zju.edu.cn}
 \affiliation{Physics Department, Zhejiang University, Hangzhou, 310027, China}
 \author{Guoping Guo}
 \email{gpguo@ustc.edu.cn}
 \affiliation{Key Laboratory of Quantum Information, University of Science and Technology of China, Hefei, 230026, China}

\begin{abstract}

In a `shortcut-to-adiabaticity' (STA) protocol, the counter-diabatic Hamiltonian, which suppresses the non-adiabatic transition
of a reference `adiabatic' trajectory, induces a quantum uncertainty of the work cost in the framework of quantum thermodynamics.
Following a theory derived recently [Funo et al 2017 {\it Phys. Rev. Lett.} {\bf118} 100602], we perform an experimental measurement of the STA work statistics
in a high-quality superconducting Xmon qubit. Through the frozen-Hamiltonian and frozen-population techniques, we
experimentally realize the two-point measurement of the work distribution for given initial eigenstates.
Our experimental statistics verify (i) the conservation of the average STA work and (ii) the equality between the STA excess of
work fluctuations and the quantum geometric tensor.

\end{abstract}

\maketitle


\section{Introduction}
\label{sec1}

The fast growing man-made quantum devices have demonstrated the potential for quantum computation, quantum information processing and quantum simulation~\cite{LaddNat10}.
Among many quantum algorithms, the utilization of an adiabatic trajectory has attracted a lot of interest in various problems~\cite{Farhi2000,FarhiSci01,Enrich16}.  
However, the bottleneck of an adiabatic quantum operation is its inevitable dissipation-induced error accumulated in a slow process.
The `shortcut-to-adiabaticity' (STA) is a theoretical protocol to speed-up the adiabatic
operation~\cite{DemirplakJPCA03,BerryJPhysA09,XChenPRL2010,MasudaPRS10,Torrontegui13AAMOPhy,CampoPRL12,CampoPRL13,TongSR15,SantosSciRep15}.
An additional counter-diabatic Hamiltonian is introduced to suppress the non-adiabatic transition and the system undergoes a
reference `adiabatic' trajectory in a short time scale. Soon after the theoretical proposal, experimental implementation has been
executed in various quantum devices~\cite{BasonNatPhy11,JFZhangPRL13,AnNatCommu16,
DuYXNatCommu16,ZhouNatPhys16,ZZXPRA17,SCPMA2018,Wangarxiv18}.
In our recent studies, we have applied the STA protocol in a superconducting qubit
for the realization of a Berry phase measurement~\cite{ZZXPRA17}, quantum state transfer~\cite{SCPMA2018}
and high-fidelity gates~\cite{Wangarxiv18}.
The STA protocol was also used by us to simulate topological phase transition             
through an experimental construction of the first Brillouin zone~\cite{SCPMA2018}.

From an energy perspective, questions about the thermodynamic nature of quantum systems have been raised~\cite{GemmerBook,SagawaBook,EspositoRMP09,CampisiRMP11},
in parallel with development of fluctuation theorems in small-scale systems~\cite{JarzynskiPRL97,CrooksPRE99}.
The statistical uncertainty of a quantum system can be separated into three categories: an initial distribution from a  pre-thermalization, a random force from a surrounding
environment, and the intrinsic uncertainty of quantum mechanics. To avoid the distinction between two basic thermodynamic variables, work and heat,
many theoretical proposals and experimental implementations have been restricted in closed systems without
quantum heat transfer between the system and the bath~\cite{HuberPRL08,DornerPRL13,MazzolaPRL13,RoncagliaPRL14,FunoPRL17,SmithNJP18,
AnNatPhys15,BatalhaoPRL14,Naghiloo2017,De18JPhysCommu,DengSciAdv18}.
To account the remaining two sources of uncertainty, the two-point measurement scheme has been proposed for
the distribution of quantum work cost~\cite{Kurchan01,Tasaki00,Mukamel03,EspositoRMP09,CampisiRMP11}. The probability of an accessible  work is a product of the probability
of a certain initial eigenstate and the  conditional probability from this initial eigenstate to an instantaneous
eigenstate at the measurement time. Despite a simple theoretical formulation, the experimental determination of the
work distribution is a nontrivial task~\cite{SmithNJP18,AnNatPhys15,BatalhaoPRL14,Naghiloo2017,De18JPhysCommu,DengSciAdv18}.

In an ideal adiabatic evolution, the intrinsic quantum uncertainty is removed as the system propagates along an adiabatic trajectory
of instantaneous eigenstates, and the work distribution is fully determined by the initial distribution. The accelerated STA
operation has been proposed to engineer friction-free quantum machines~\cite{DengPRE13,CampoSciRep14,BeauEntrpy16}.
However, the exact instantaneous eigenbasis is rotated away from the reference adiabatic basis due to the introduction of the counter-diabatic
Hamiltonian. Therefore, each STA protocol is physically characterized by its deviation of the work statistics, which is originated from
the quantum uncertainty of the counter-diabatic Hamiltonian. In a recent paper~\cite{FunoPRL17}, the statistics of the STA work were
theoretically resolved. An equality connecting the STA excess of work fluctuations and the quantum geometric tensor was derived.
The experimental verification of this STA work theory was proposed for a harmonic oscillator in a trapped ion system.
Alternatively, high-quality qubits with sufficiently long quantum coherence have been achieved in
a superconducting circuit~\cite{ChowPRL09,BarendsPRL13,BarendsNat14,KellyNat15,YanNatCommu16,KandalaNat17}.
Sophisticated microwave control techniques allow us to precisely manipulate and measure a superconducting qubit system.
In this paper, we drive a single superconducting Xmon qubit with an STA field and experimentally determine the statistics of
the STA work for a given initial eigenstate. To realize the two-point measurement scheme, we design the frozen-Hamitlonian and frozen-population sequences
to extract the instantaneous eigenenergies and the population of instantaneous eigenstates, respectively. Our study thus
reports an experimental verification of the STA work theory in Ref.~\cite{FunoPRL17}.


\section{Theory}
\label{sec2}

In this section, we will briefly review the theory of STA work fluctuations in Ref.~\cite{FunoPRL17} and apply it
to the quantum system of a single qubit.

\subsection{STA Protocol}
\label{sec2a}

For an arbitrary non-degenerate quantum system driven by a time-dependent Hamiltonian $H_0(t)$,
we introduce its instantaneous eigenbasis $\{\left|n(t)\right\rangle\}$ satisfying $H_0(t)|n(t)\rangle=\varepsilon_n(t) |n(t)\rangle$,
where $\varepsilon_n(t)$ is the $n$-th instantaneous eigenenergy and  $\left|n(t)\right\rangle$ is
the associated eigenstate. The Hamiltonian is thus expanded in this instantaneous eigenbasis,
becoming
\be
H_0(t)=\sum_n\varepsilon_n(t) |n(t)\rangle\langle n(t)|.
\label{eq_01}
\ee
In the scenario of $d_t H_0(t)\rightarrow0$, the system follows an adiabatic trajectory, $|n(0)\rangle\rightarrow|n(t)\rangle$,
if it is  prepared  at $|n(0)\rangle$ initially. The quantum state at time $t$ is given by $|\Psi(t)\rangle=U_\ad(t)|n(0)\rangle=c_n(t)|n(t)\rangle$,
where $U_\ad(t) = T_+ \exp[-(i/\hbar)\int_0^t H_0(\tau)d\tau]$ is the adiabatic time evolution operator
and the coefficient $c_n(t)$ includes the accumulation of both dynamic and geometric phases~\cite{Berry1984}.
Here $T_+$ denotes the time ordering operator and $\hbar=h/2\pi$ is the reduced Planck constant.

In practice, we often hope that the quantum system can follow an adiabatic trajectory but
in a short time scale. In the STA protocol~\cite{DemirplakJPCA03,BerryJPhysA09,XChenPRL2010,MasudaPRS10,Torrontegui13AAMOPhy,CampoPRL12,CampoPRL13,TongSR15,SantosSciRep15},
an additional counter-diabatic Hamiltonian $H_\cd(t)$ is constructed to cancel the non-adiabatic transition.
In a formally exact way, $H_\cd(t)$  is written as~\cite{BerryJPhysA09}
\be
H_\cd(t) &=& i\hbar\sum_n \calP^\perp_{n}(t) |\partial_t n(t)\rangle\langle n(t)|, 
\label{eq_02}
\ee
where $\calP^\perp_{n}(t) = 1-| n(t)\rangle\langle n(t)|$ is the projection operator onto the subspace perpendicular to $|n(t)\rangle$.
The quantum system driven by $H(t)=H_0(t)+H_\cd(t)$  evolves rigorously along the adiabatic trajectory of $H_0(t)$,
i.e., $|\Psi(t)\rangle=U_\STA(t)|n(0)\rangle=c_n(t)|n(t)\rangle$
with the STA time evolution operator $U_\STA(t) = T_+\exp[-(i/\hbar)\int_0^t H(\tau)d\tau]$.

In general, the reference Hamiltonian $H_0(t)$ may be expressed as a function of
time-dependent control parameters, $\lambda(t) = \{\lambda_1(t), \lambda_2(t), \cdots\}$~\cite{Berry1984}.
A formal representation, $H_0(t) \equiv H_0(\lambda)$, is applied, where the $t$-dependence of $\lambda$
is implicitly assumed. The same representation can be applied to the reference instantaneous eigenstates,
i.e. $|n(t)\rangle \equiv |n(\lambda)\rangle$.
Based on its definition in Eq.~(\ref{eq_02}), the counter-diabatic Hamiltonian
is formulated as $H_\cd(t)\equiv H_\cd(\lambda, \dot{\lambda})$ with
\be
H_\cd(\lambda, \dot{\lambda}) = i \hbar \sum_n \sum_\mu \calP^\perp_{n}(\lambda) \left|\partial_{\mu} n(\lambda)\right\rangle\left\langle n(\lambda)\right| \dot{\lambda}_\mu.
\label{eq_02a}
\ee
The symbol $\partial_{\mu}$ stands for $\partial/\partial \lambda_\mu$
and the time derivatives, $\dot{\lambda}(t)=\{\dot{\lambda}_1(t), \dot{\lambda}_2(t), \cdots\}$,
specify a designed  STA protocol.

In our experiment, we study a single superconducting Xmon qubit
which can be mapped onto a spin-$1/2$ particle
driven by an external field~\cite{ChuangBook}. For consistency, we will mostly take the notation
of the up ($\left|\uparrow\right\rangle$) and down ($\left|\downarrow\right\rangle$)
states rather than the equivalent ground ($|0\rangle$) and excited ($|1\rangle$) states.
In the rotating frame, the time-dependent reference Hamiltonian  follows a general form,
\be
H_0(t) &=& \hbar \bm B_0(t)\cdot\bm \sigma/2,
\label{eq_03}
\ee
where $\bm B_0(t)\!=\!\Omega(t)(\sin\theta(t)\cos\phi(t),\sin\theta(t)\sin\phi(t),\cos\theta(t))$
is the vector of an external field and  ${\boldsymbol \sigma}\!=\!(\sigma_x, \sigma_y, \sigma_z)$ is the vector of Pauli matrices.  
The time-dependent control parameters  are the amplitude $\Omega(t)$, the polar angle $\theta(t)$ and the azimuthal angle $\phi(t)$,
i.e., $\lambda(t) = \{\Omega(t), \theta(t), \phi(t)\}$.
By calculating the reference instantaneous eigenstates,
\be
\left\{\ba{lll}
|s_\uparrow(t)\rangle &=& \cos[\theta(t)/2] \left|\uparrow\right\rangle + \sin[\theta(t)/2]e^{i\phi(t)} \left|\downarrow\right\rangle \\
|s_\downarrow(t)\rangle &=& -\sin[\theta(t)/2]e^{-i\phi(t)}\left|\uparrow\right\rangle + \cos[\theta(t)/2] \left|\downarrow\right\rangle \ea \right.
\label{eq_03a}
\ee
and substituting them into Eq.~(\ref{eq_02}), we obtain an analytical expression of the counter-diabatic Hamitlonian, which reads
\be
H_\cd(t)=\hbar \bm B_\cd(t)\cdot\bm \sigma/2.
\label{eq_04}
\ee
The three elements of the counter-diabatic field $\bm B_\cd(t)\!=\!(B_{\cd; x}(t), B_{\cd; y}(t), B_{\cd; z}(t))$
are explicitly given by~\cite{DemirplakJPCA03,BerryJPhysA09,ZZXPRA17,SCPMA2018,Wangarxiv18}
\be
\left\{ \ba{ccl} B_{\cd; x}(t) &=&   -\dot{\theta}(t)\sin\phi(t)-\dot{\phi}(t)\sin\theta(t)\cos\theta(t)\cos\phi(t) \\
B_{\cd; y}(t) &=&  \dot{\theta}(t)\cos\phi(t)-\dot{\phi}(t)\sin\theta(t)\cos\theta(t)\sin\phi(t) \\
B_{\cd; z}(t) &=&  \dot{\phi}(t)\sin^2\theta(t) \ea\right. .
\label{eq_05}
\ee
The total STA field is a sum of the reference and counter-diabatic fields, i.e., $\bm B(t)=\bm B_0(t)+\bm B_{\rm cd}(t)$.

\subsection{Statistics of STA Work}
\label{sec2b}

Although the non-adiabatic transition is fully suppressed in the STA protocol,
the introduction of the counter-diabatic Hamiltonian $H_\cd(t)$ is not cost-free.
For the total Hamiltonian $H(t)$,  we introduce the STA instantaneous eigenbasis $\{\left|\psi_k(t)\right\rangle \}$, which leads to
\be
H(t) = \sum_k E_k(t) |\psi_k(t)\rangle\langle \psi_k(t)|
\label{eq_07}
\ee
with $E_k(t)$ the $k$-th instantaneous eigenenergy and $|\psi_k(t)\rangle$ the associated eigenstate. 
For a given initial state of $\left|\Psi(0)\right\rangle =|n(0)\rangle$, the probability
of observing the quantum state $|\psi_k(t)\rangle$ at time $t$  is given by
\be
P_{k|n}(t) = |\langle\psi_k(t)|U_\STA(t)|n(0)\rangle|^2 = |\langle\psi_k(t)|n(t)\rangle|^2.
\label{eq_08}
\ee
The corresponding joint probability is written as $P_{k n}(t)=P_{k|n}(t) P_n(0)$ with $P_n(0)$ an initial probability at
the state $|n(0)\rangle$.

Next we introduce the concept of an STA work in the framework of quantum thermodynamics.
The initial system is assumed to follow a canonical distribution,
$\allowbreak \rho(0)= \sum_n P_n(0) |n(0)\rangle\langle n(0)|$ with $P_n(0)\propto \exp[-\bt \varepsilon_n(0)]$.
Based on the two-point measurement scheme~\cite{Kurchan01,Tasaki00,CampisiRMP11,Mukamel03}, the probability distribution function of
a quantum work at time $t$ is written as
\be
P(W, t) &=& \sum_{nk}\delta(W-\delta E_{kn}(t))P_{kn}(t)
\label{eq_09}
\ee
with $\delta E_{kn}(t)= E_k(t)-\varepsilon_n(0)$. In Eq.~(\ref{eq_09}), we implicitly assume
that the initial counter-diabatic Hamiltonian is zero, i.e., $H_\cd(0)=0$~\cite{FunoPRL17}.
The general $m$-th moment of the quantum work is defined as
\be
\langle W^m(t)\rangle &=& \int W^m P(W, t)dW,
\label{eq_10}
\ee
which can be rewritten as
\be
\langle W^m(t)\rangle = \sum_n \overline{W^m_n(t)}P_n(0).
\label{eq_11}
\ee
The quantity, $\overline{W^m_n(t)} = \sum_k [E_k(t)-\varepsilon_n(0)]^m P_{k|n}(t)$,
is the $m$-th moment of the quantum work for the initial eigenstate
$|n(0)\rangle$. This term is independent of the
initial distribution but fully determined by the designed STA protocol~\cite{FunoPRL17}.
In this paper, we will present an experimental investigation of $\overline{W^m_n(t)}$ instead of
the statistics of the total work $\langle W^m(t)\rangle$.

In an adiabatic process, the conditional probabilities satisfy $P_{k|n}(t)=\delta_{k, n}$ and
the $m$-th moment is simplified to be $\overline{W^m_{\ad; n}(t)} = [\varepsilon_n(t)-\varepsilon_n(0)]^m$.
In the STA protocol, this quantity is changed 
due to an quantum uncertainty induced by $H_\cd(t)$.
Here we focus on the first and second moments, which are related to the average work and its variance.
Through a straightforward derivation (see Appendix~\ref{app1}), we obtain the first equality~\cite{FunoPRL17},
\be
\overline{W_{n}(t)} &=& \overline{W_{\ad; n}(t)} + i\hbar \langle n(t) |\calP^\perp_n |\partial_t n(t)\rangle.
\label{eq_13}
\ee
The second term on the right-hand side (RHS) of Eq.~(\ref{eq_13}) vanishes due to an orthogonal projection, 
$\langle n(t)| \calP^\perp_n=0$.
In the STA protocol,  the preservation of the adiabatic trajectory, $|n(0)\rangle\rightarrow |n(t)\rangle$,
is represented alternatively by the conservation of the average work, i.e.,
\be
\overline{W_{n}(t)} = \overline{W_{\ad; n}(t)}.
\label{eq_13a}
\ee
However, the quantum uncertainty created in the STA instantaneous eigenbasis $\{|\psi_k(t)\rangle\}$
cannot be cancelled in the second order moment of the STA work.
Through another straightforward derivation (see Appendix~\ref{app1}), we obtain the second equality~\cite{FunoPRL17},
\be
\overline{W^2_n(t)} &=& \overline{W^2_{\ad; n}(t)} + \hbar^2 \langle \partial_t n(t) |\calP^\perp_n |\partial_t n(t)\rangle.
\label{eq_14}
\ee
With the introduction of the control parameter set $\lambda(t)$, Eq.~(\ref{eq_14}) is rewritten as
\be
\delta\overline{ W^2_{n}(t)} = \hbar^2 \sum_{\mu,\nu} g^{(n)}_{\mu\nu} \dot{\lambda}_\mu\dot{\lambda}_\nu,
\label{eq_15}
\ee
where $\delta\overline{ W^2_{n}(t)}=\overline{W^2_n(t)}-\overline{W^2_{\ad; n}(t)}$
denotes the STA  excess of work fluctuations. The term,
$g^{(n)}_{\mu\nu}={\rm Re}\,  Q^{(n)}_{\mu\nu} = {\rm Re}\, \langle \partial_\mu n(\lambda) |\calP^\perp_n |\partial_\nu n(\lambda)\rangle$,
is the real part of the quantum geometric tensor of
the $|n(t)\rangle$-state manifold defined in the parameter space of $\lambda$~\cite{FunoPRL17,Provost1980}.
The time dependence on the RHS of Eq.~(\ref{eq_15}) arises from the varying speed of the control parameters.
Since the eigenvalues of the  $g^{(n)}_{\mu\nu}$-tensor are always equal or greater than zero, the STA work variance
is always equal or greater than that under the adiabatic condition, i.e.,
\be
\overline{W^2_n(t)} \ge \overline{W^2_{\ad; n}(t)}.
\label{eq_15a}
\ee

For our spin-1/2 particle, the instantaneous eigenenergies
of the STA Hamiltonian $H(t)$ are $E_\pm(t)=\pm \hbar |\bm B(t)|/2$ and the associated eigenstates
are denoted as $|\psi_\pm (t)\rangle$. Accordingly, the four conditional probabilities involved
are $P_{\pm | \uparrow}(t)=|\langle\psi_\pm(t)|U_\STA(t) |s_\uparrow(0)\rangle^2$ and $P_{\pm |\downarrow}(t)=\langle \psi_\pm(t) |U_\STA(t)|s_\downarrow(0)\rangle^2$.
Since the reference instantaneous eigenstates 
are independent of the amplitude $\Omega(t)$,
a two-element set, $\lambda(t)=\{\theta(t), \phi(t)\}$, is used to calculate the geometric tensor~\cite{FunoPRL17,Provost1980}, 
\be
g^{\uparrow}(\lambda)=g^\downarrow(\lambda)=\frac{1}{4}
\left(\ba{cc}
1    &0   \\
0& \sin^2\theta \ea \right).
\label{eq_16}
\ee
In our experiment, the STA protocol is designed to compress a target adiabatic process evenly through time.
With an operation time $T$, the reference field is defined by the same form of $\bm B_0(\tilde{t}=t/T)$.
Consequently, we obtain the STA excess of work fluctuations in the
single-qubit system as
\be
\delta\overline{ W^2_{n}(t)} &=& \frac{\hbar^2}{4 T^2} \left[\left(\frac{d\theta}{d\tilde{t}}\right)^2+\sin^2\theta\left(\frac{d\phi}{d\tilde{t}}\right)^2 \right],
\label{eq_17}
\ee
which shows an inverse square dependence of the STA operation time $T$.

\begin{figure}[tp]
\centering
\includegraphics[width=0.65\columnwidth]{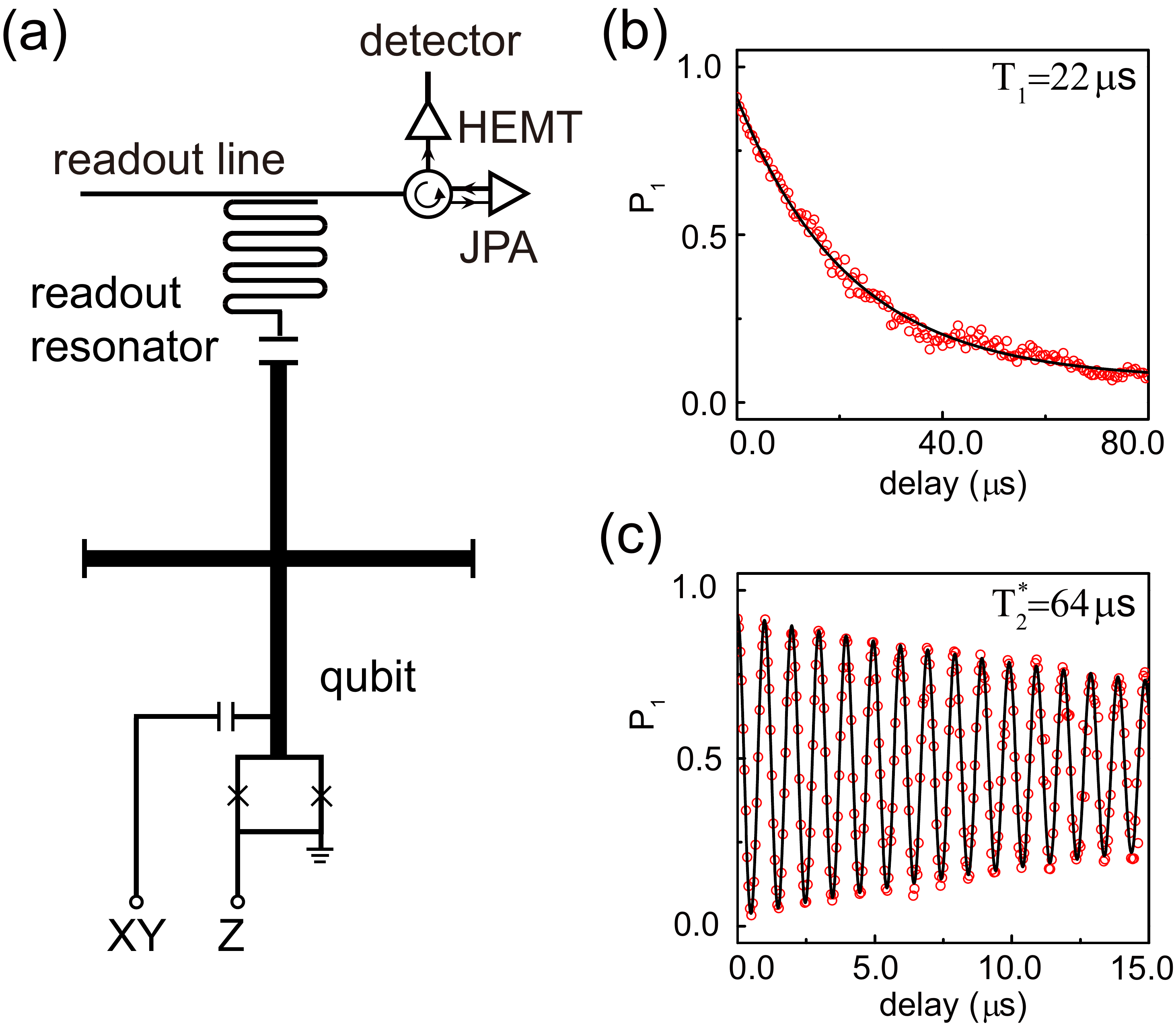}
\caption{(a) A schematic diagram of a single cross-shaped Xmon qubit. (b) Population relaxation of the Xmon qubit is shown in red circles, fitted by a solid line with a relaxation time of $T_1=22$ $\mu$s.
(c) Ramsey fringes of the Xmon qubit are shown in red circles, fitted by a solid line with $T_1$ and a pure decoherence time of $T^*_2=64$ $\mu$s.}
\label{fig_n01}
\end{figure}

\section{Experimental Setup}
\label{sec3}

Our study of the STA work is performed in a cross-shaped Xmon qubit~\cite{BarendsPRL13,BarendsNat14,KellyNat15},
with the same experimental setup as in Ref.~\cite{Wangarxiv18}.
Through the bottom arm of the cross (see Figs.~\ref{fig_n01}(a)), the qubit is dc-biased at a resonance frequency
of $\omega_{10}/2\pi=4.85$ GHz, where $\hbar\omega_{10}$
is the energy difference between the ground  ($|{0}\rangle$) and excited  ($|{1}\rangle$) states.
Also through the bottom arm, a microwave signal is generated separately to drive the qubit to undergo
a designed passage. 
In the dispersive readout, the qubit state is encoded in the frequency shift of the readout resonator
which is coupled to the top arm of the cross. The readout signal is sent through a
transmission line and a series of circulators, further being amplified by a Josephson parametric amplifier (JPA)
and a high electron mobility transistor (HEMT) for a high fidelity measurement~\cite{RoyAPL15,XiaoYuanPRL16}.
Compared with our previous experiment~\cite{Wangarxiv18}, the qubit coherence at the sweet point
is further improved with a relaxation time of $T_1=22$ $\mu$s and a pure decoherence time
of $T_2^{*}=64$ $\mu$s (see Figs.~\ref{fig_n01}(b) and~\ref{fig_n01}(c)).
The qubit device is measured in a dilution refrigerator whose base temperature is $\sim 10$ mK.

\section{Results}
\label{sec4}

\begin{figure}[tp]
\centering
\includegraphics[width=0.65\columnwidth]{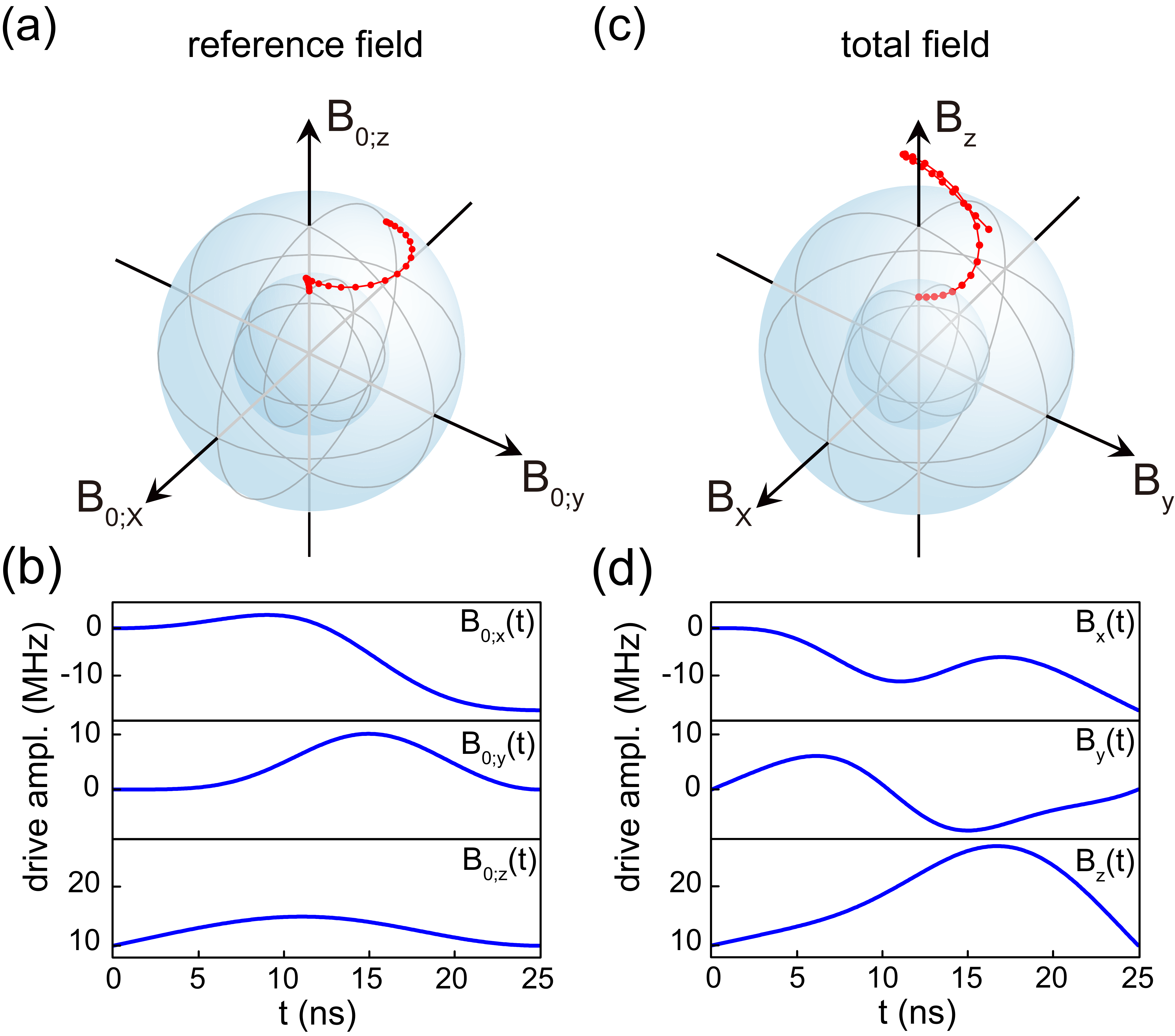}
\caption{(a-b) The reference adiabatic field $\bm B_0(t)$: (a) its trajectory in the parameter space and (b) the
 time evolution of the three components along $x$, $y$ and $z$ directions. (c-d) The total field $\bm B(t)$ that is
 the sum  of the reference and counter-diabatic fields: (c) its trajectory in the parameter space and (d) the
 time evolution of the three components. The operation time is set at $T=25$ ns. }
\label{fig_n02}
\end{figure}

In our experiment, the reference adiabatic field ${\boldsymbol B}_0(t) = \Omega(\sin\theta\cos\phi,\sin\theta\sin\phi,\cos\theta)$ is designed
as
\be
\left\{\ba{lll}
\Omega(t) &=& \Omega_0 + \Omega_1 \sin(\pi t/2T)\\
\theta(t) &=& (\pi/6) [ 1-\cos(\pi t/T)]\\
\phi(t) &=& (\pi/2) [ 1- \cos(\pi t/T)]
\ea
\right. .
\label{eq_18}
\ee
The two amplitude parameters are set at $\Omega_0/{2\pi}= 10$ MHz and $\Omega_1/{2\pi} = 10$ MHz.
As shown in Fig.~\ref{fig_n02}(a), this reference field $\bm B_0(t)$ points to the $z$-axis initially
so that the initial eigenstates of $H_0(0)$ are given by $|s_\uparrow(0)\rangle=\left|\uparrow\right\rangle(=\left|0\right\rangle)$
and $|s_\downarrow(0)\rangle=\left|\downarrow\right\rangle(=\left|1\right\rangle)$.
To demonstrate nontrivial geometric tensors, the reference adiabatic trajectory is not geodesic.
Through the evolution of $\bm B_0(t)$,  the amplitude is doubled, the polar angle is rotated by 60$^\circ$,
and the azimuthal angle is rotated by 180$^\circ$. In our experiment, the influence of higher excited states is very small
with the design of $\bm B_0(t)$ in Eq.~(\ref{eq_18})~\cite{SCPMA2018,Wangarxiv18}.
With an example operation time, $T=25$ ns,
the three components of $\bm B_0(t)$ along the $x$-, $y$- and $z$-directions are
plotted in Fig.~\ref{fig_n02}(b). In response to this reference field,
the Bloch vector of the reference instantaneous eigenstate $\left|s_\uparrow(t)\right\rangle$
experiences the same rotations along the polar and azimuthal angles as $\bm B_0(t)$,
while the Bloch vector of $\left|s_\downarrow(t)\right\rangle$ experiences the reversed rotations.
The change of the amplitude $\Omega(t)$ induces an adiabatic work cost for both $\left|s_\uparrow(t)\right\rangle$
and $\left|s_\downarrow(t)\right\rangle$. Subsequently, the counter-diabatic field $\bm{B}_\cd(t)=(B_{\cd; x}(t), B_{\cd; y}(t), B_{\cd; z}(t))$
is analytically obtained as
\be
\left\{\ba{lll}
B_{\cd; x}(t) &=& -(\pi^2/6T)\sin(\pi t/T)\left[\sin\phi(t)+3\sin\theta(t)\cos\theta(t)\cos\phi(t) \right]\\
B_{\cd; y}(t) &=& (\pi^2/6T)\sin(\pi t/T)\left[\cos\phi(t)-3\sin\theta(t)\cos\theta(t)\sin\phi(t) \right]\\
B_{\cd; z}(t) &=& (\pi^2/2T)\sin(\pi t/T)\sin^2\theta(t)
\ea
\right. .
\label{eq_19}
\ee
The total STA field supplied in our experiment is given by $\bm B(t)=\bm B_0(t)+\bm{B}_\cd(t)$.
Equation~(\ref{eq_19}) shows that the counter-diabatic field vanishes at the initial
and final moments, i.e., $\bm B_{\rm cd}(0)=0$ and $\bm B_{\rm cd}(T)=0$.
The initial STA eigenbasis
$\{\left|\psi_+(0)\right\rangle, \left|\psi_-(0)\right\rangle\}$ is identical to
the adiabatic eigenbasis $\{\left|s_\uparrow(0)\right\rangle,\left|s_\downarrow(0)\right\rangle\}$,
which satisfies the presumption of the two-point measurement scheme~\cite{Kurchan01,Tasaki00,Mukamel03,CampisiRMP11}.
In Figs.~\ref{fig_n02}(c)
and~\ref{fig_n02}(d), the evolution of $\bm B(t)$ with the operation time $T=25$ ns is
plotted explicitly.  In the intermediate time range (especially around $t\approx 16$ ns),
a large difference is found between $\bm B_0(t)$ and $\bm B(t)$, due to the acceleration of adiabaticity in a short operation time.
With the increase of $T$, the amplitude of the counter-diabatic field is decreased and
the adiabatic limit of $\bm B(t)\approx \bm B_0(t)$ is gradually approached.

\subsection{Frozen-Hamiltonian and Frozen-Population Measurements in the Instantaneous  Eigenbasis}
\label{sec4a}

The experimental verification of STA work fluctuations requires a measurement scheme
in the instantaneous eigenbasis of the total Hamiltonian $H(t)$.
Here we design two different sequences to detect the  eigenenergies and the population of eigenstates separately.

\begin{figure}[tp]
\centering
\includegraphics[width=0.65\columnwidth]{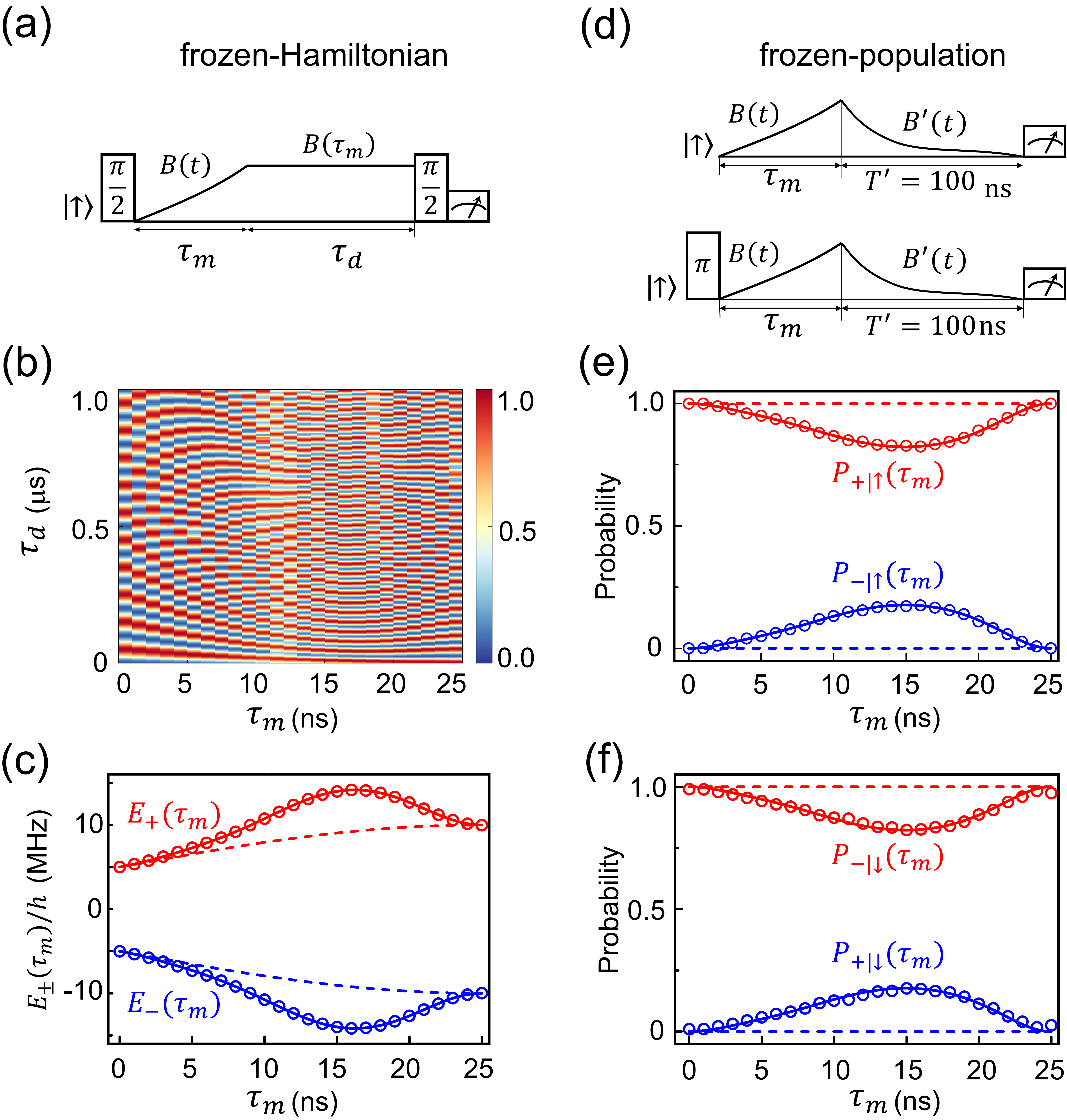}
\caption{(a-c) Measurement of the STA instantaneous eigenenergies. (a) A frozen-Hamiltonian scheme to measure $E_\pm(\tau_m)$ at an intermediate time $\tau_m$ (see text for details).
(b) The up-state population $P_\uparrow(\tau_m+\tau_d)$ with the change of the interruption time $\tau_m$ and the extra duration time $\tau_d$.
For each $\tau_m$, $P_\uparrow(\tau_m+\tau_d)$ oscillates with $\tau_d$, which leads to (c) the experimental determination of $E_{\pm}(\tau_m)$.
(d-f) Measurement of the population at STA instantaneous eigenstates. (d) A frozen-population scheme to measure $P_{\pm|\uparrow}(\tau_m)$ in the upper panel
and $P_{\pm|\downarrow}(\tau_m)$ in the lower panel (see text for details).  (e) and (f) present the experimental measurement of $P_{\pm|\uparrow}(\tau_m)$
and $P_{\pm|\downarrow}(\tau_m)$, respectively. In (c), (e) and (f), the open circles are experimental results, while the dashed and  solid lines
are theoretical predictions under the reference adiabatic and STA operations, respectively. Here the STA operation time is set at $T=25$ ns. }
\label{fig_n03}
\end{figure}

The eigenenergy measurement scheme is shown in Fig.~\ref{fig_n03}(a). A $\pi/2$-pulse is initially applied to the
up state (equivalent to the ground state) to create a superposition state $(\left|\uparrow\right\rangle+\left|\downarrow\right\rangle)/\sqrt{2}$,
which is then driven by the STA field $\bm{B}(t)$. At an intermediate time $\tau_m$ ($0\!\le\!\tau_m\!\le\! T$), the external field
is frozen for an extra time duration of $\tau_d$, giving $\bm{B}(\tau_m\!\le\! t\! \le\!\tau_m+\tau_d)\!=\!\bm{B}(\tau_m)$. Since the quantum state 
at time $\tau_m$ is a linear combination of two instantaneous eigenstates, $|\Psi(\tau_m)\rangle=c_+(\tau_m)|\psi_+(\tau_m)\rangle+c_-(\tau_m)|\psi_-(\tau_m)\rangle$,
a quantum oscillation is expected in the frozen period, where
the oscillation frequency is determined by $E_+(\tau_m)\!-\!E_-(\tau_m) = 2E_+(\tau_m)$.
This frozen-Hamiltonian measurement can experimentally detect the instantaneous eigenenergies,
$E_+(\tau_m)$ and $E_-(\tau_m)$~\cite{RoushanSci17}. In Fig.~\ref{fig_n03}(a),
a Ramsey-type sequence is used to enhance the amplitude of quantum oscillations despite the fact that an arbitrary initial state is
allowed. For $T=25$ ns, Fig.~\ref{fig_n03}(b) presents the experimental measurement
of the up-state population $P_\uparrow(\tau_m\!+\!\tau_d)$ with the change of the interruption
time $\tau_m$ and the duration time $\tau_d$. The time interval of $\tau_m$ is 1 ns and a total time range of 1 $\mu$s
is swept for $\tau_d$. For each value of $\tau_m$, the measured population $P_\uparrow(\tau_m\!+\!\tau_d)$
periodically oscillates with $\tau_d$ and the oscillation frequency varies with $\tau_m$.
The experimentally extracted results of $E_\pm(\tau_m)$ are plotted in Fig.~\ref{fig_n03}(c).
An excellent agreement is observed between the experimental measurement and the theoretical design.
Compared to the adiabatic increase of $\varepsilon_{\uparrow}(\tau_m)/h$ from 5 MHz to 10 MHz,
the STA protocol with $T=25$ ns induces a bump of 8 MHz around $\tau_m\approx 16$ ns for the instantaneous eigenenergy $E_+(\tau_m)/h$,
where $h$ is the Planck constant.

Figure~\ref{fig_n03}(d) demonstrates the scheme of measuring the population of the two STA instantaneous eigenstates.
The qubit is initialized at the up state that is
an initial eigenstate, i.e., $|s_\uparrow(0)\rangle=\left|\uparrow\right\rangle$, and then driven by the STA field $\bm{B}(t)$.
At each intermediate time $\tau_m$,
the external field $\bm{B}(t)$ is suddenly interrupted and replaced by another field $\bm B^\pr_0(t)$,
which freezes the population, $P_{\pm|\uparrow}(\tau_m)=|c_\pm(\tau_m)|^2$, by dragging the qubit adiabatically.
If the final external field is ramped back to the $z$-direction, the population measured at the up
and down states provides an experimental determination of $P_{\pm|\uparrow}(\tau_m)$~\cite{SmithNJP18}.
To avoid a non-adiabatic error, the STA protocol is used pratically with $\bm B^\pr(t)=\bm B^\pr_0(t)+\bm B^\pr_{\rm cd}(t)$.
A great flexibility is allowed in the functional form of $\bm B^\pr_0(t)$. In our experiment,
we choose a geodesic line for the reference adiabatic trajectory of $\bm B^\pr_0(t)/|\bm B^\pr_0(t)|$
and this second STA operation time is set at $T^\pr=100$ ns.
The measured result of $P_{\pm|\uparrow}(\tau_m)$ for $T=25$ ns is plotted  in Fig.~\ref{fig_n03}(e),
showing the same excellent agreement with the theoretical prediction.
The quantum uncertainty, $P_{+|\uparrow}(\tau_m)\!<\!1$ and $P_{-|\uparrow}(\tau_m)\!>\!0$, in the STA operation
is observed through time. The maximum uncertainty of $\sim 20\%$ appears around $\tau_m\approx16$ ns.
We perform the experimental measurement of $P_{\pm|\downarrow}(\tau_m)$ by applying a $\pi$-pulse to
the up-state qubit followed by the same frozen-population approach (see Fig.~\ref{fig_n03}(d)).
Since the down state is equivalent to the excited state of the qubit,
an error is accumulated in the experimental measurement of $P_{\pm|\downarrow}(\tau_m)$
due to inevitable quantum dissipation of $T_1$ and $T_2=[(2T)^{-1}_1+(T^\ast_2)^{-1}]^{-1}$ (see Fig.~\ref{fig_n03}(f)).
This dissipation-induced error is gradually increased with the increase of the operation time $T$.

\begin{figure}[tp]
\centering
\includegraphics[width=0.65\columnwidth]{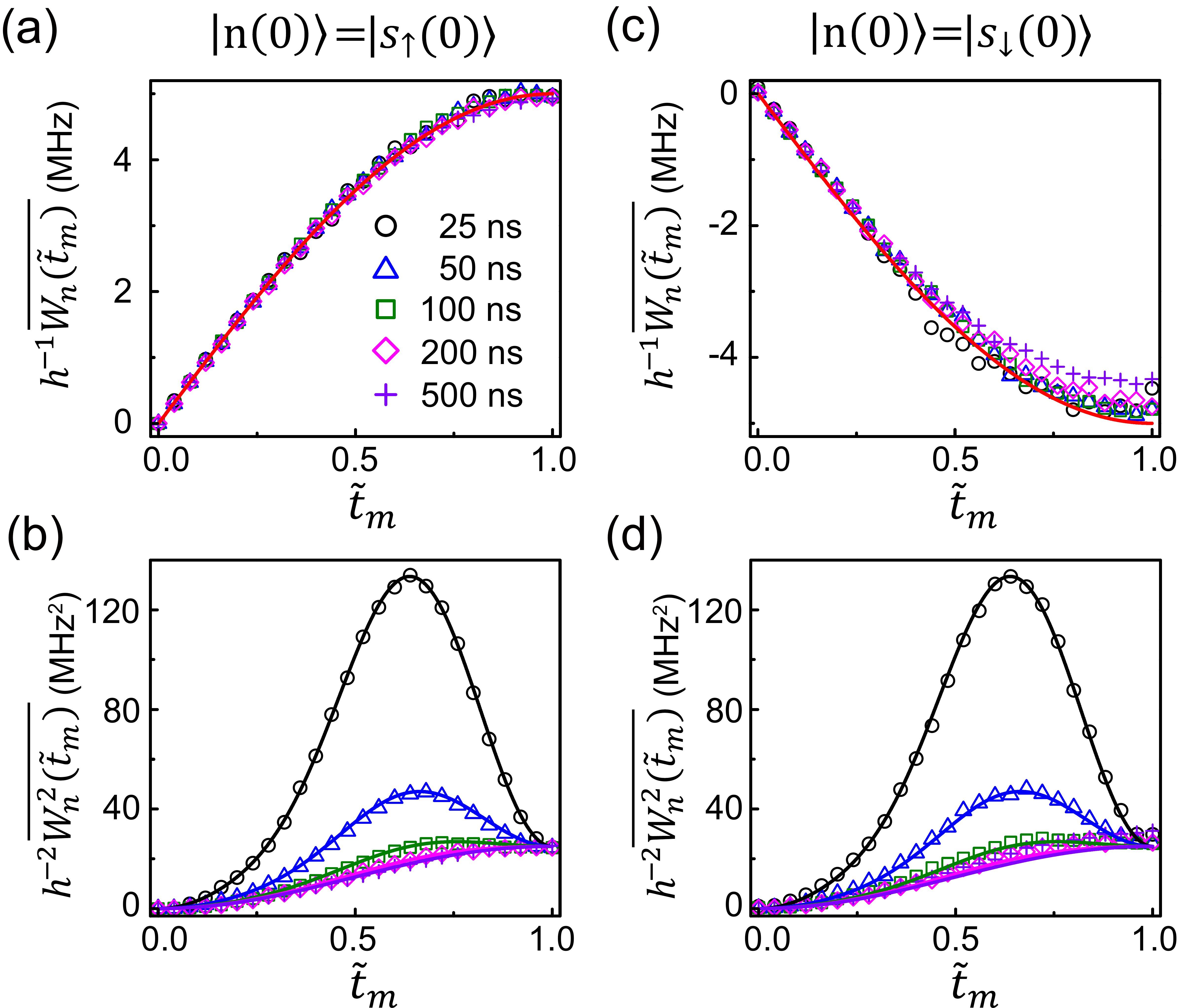}
\caption{Experimental statistics of STA work fluctuations. For the initial eigenstate $|n(0)\rangle=|s_\uparrow(0)\rangle$
the first and second moments, $\overline{W_\uparrow(\tilde{t}_m)}$ and $\overline{W^2_\uparrow(\tilde{t}_m)}$, are shown in (a) and (b), respectively.
For the initial eigenstate $|n(0)\rangle=|s_\downarrow(0)\rangle$, the first and second moments,
$\overline{W_\downarrow(\tilde{t}_m)}$ and $\overline{W^2_\downarrow(\tilde{t}_m)}$,
are shown in (c) and (d), respectively. The total 5 operation times,  $T=25$, 50, 100, 200 and 500 ns, are experimentally studied.
The measurement time is rescaled by $\tilde{t}_m=\tau_m/T$ for each operation time. The experimental results are shown in symbols
while the theoretical predictions are shown in solid lines.}
\label{fig_n04}
\end{figure}

\subsection{The First and Second Moments of STA Work}
\label{sec4b}

The measurement of the STA instantaneous eigenenergies, $E_\pm(\tau_m)$,
and the conditional probabilities, $P_{\pm|\uparrow}(\tau_m)$ and $P_{\pm|\downarrow}(\tau_m)$,
allows us to obtain experimental statistics of the STA work.
In our experiment, the STA operation time is set at $T=25$, 50, 100, 200 and 500 ns.
The STA evolution with $T=500$ ns is very close to  the adiabatic passage since $\bm B_{\rm cd}(t)$
is nearly negligible.

The first moment of the STA work at each measurement time $\tau_m$ is experimentally determined by $\overline{W_{n}(\tau_m)}=\sum_k [E_k(\tau_m)-\varepsilon_n(0) ]P_{k|n}(\tau_m)$
for both initial conditions, $|n(0)\rangle\!=\!\left|s_\uparrow(0)\right\rangle$ and $\left|s_\downarrow(0)\right\rangle$.
For a given operation time $T$, we rescale the measurement time by $\tilde{t}_m\!=\!\tau_m/T$  and present the result of
$\overline{W_{n}(\tilde{t}_m)}$ as a function of the reduced time $\tilde{t}_m$.
Since the reference external field with different values of $T$ follows the same form of $\bm B_0(\tilde{t}=t/T)$,
the rescaled time provides a transparent comparison on the influence of the STA
operation time. As shown in Figs.~\ref{fig_n04}(a) and~\ref{fig_n04}(b),
the data of $\overline{W_{n}(\tilde{t}_m)}$ with different $T$ collapse to a single curve for each initial condition,
which is consistent with the theoretical prediction in Eq.~(\ref{eq_13a})~\cite{FunoPRL17}. With the designed reference  
field in Eq.~(\ref{eq_18}), the averaged STA work follows the adiabatic behavior of
$\overline{W_{n}(\tilde{t}_m)}=\overline{W_{\ad; n}(\tilde{t}_m)}=\pm [\Omega(\tilde{t}_m)-\Omega(0)]/2$. A systematic error
is found for $\overline{W_{\downarrow}(\tilde{t}_m)}$, especially with $T=500$ ns, which is due to the
quantum dissipation of the qubit (see Sec.~\ref{sec4a}).

The second moment of the STA work is experimentally determined by $\overline{W^2_{n}(\tau_m)}=\sum_k [E_k(\tau_m)-\varepsilon_n(0) ]^2P_{k|n}(\tau_m)$.
The experimental results of $\overline{W^2_{\uparrow}(\tilde{t}_m)}$ and $\overline{W^2_{\downarrow}(\tilde{t}_m)}$ are
shown in Figs.~\ref{fig_n04}(c) and~\ref{fig_n04}(d), respectively. Different from the behavior of the averaged STA work,
the work variance  increases monotonically with the decrease of the STA operation time. Since the STA operation
with $T=500$ ns can mimic the adiabatic process, we thus experimentally verifies the inequality of the STA
work fluctuations, $\overline{W^2_{n}(\tilde{t}_m)}\ge\overline{W^2_{\ad; n}(\tilde{t}_m)}$~\cite{FunoPRL17}. The equality is obtained
at the initial and final moments where the counter-diabatic field  vanishes, given by $\bm B_\cd(0)=\bm B_\cd(T)=0$. As a comparison,
both $\overline{W^2_{\uparrow}(\tilde{t}_m)}$ and $\overline{W^2_{\downarrow}(\tilde{t}_m)}$ agree excellently
with the theoretical prediction for $T=25$, 50 and 100 ns. As the operation time is further increased to
$T=200$ and 500 ns, the dissipation-induced error deviates the experimental result of $\overline{W^2_{\downarrow}(\tilde{t}_m)}$
away from the theoretical prediction.


\subsection{Experimental Verification of the Relation between the STA Work Fluctuations and the Quantum Geometric Tensor}
\label{sec4c}

Based on the theoretical design, the STA excess of work fluctuations,
$\delta\overline{W^2_n(t)}=\overline{W^2_n(t)}-\overline{W^2_{\ad; n}(t)}$, is fully determined by the functional form
of $H_\cd(t)$. From an alternative perspective of differential geometry, $H_\cd(t)$ guides
a parallel transport of the reference instantaneous eigenstate $|n(t)\rangle$ by satisfying
$\langle n(t)| H_\cd(t) |  n(t)\rangle\!=\!0$~\cite{FunoPRL17,AnandanPhysLettA88}.
With a geometric tensor,
$g^{(n)}_{\mu\nu}=\mathrm{Re}\!\left\langle \partial_\mu n(\lambda) | P^\perp_n | \partial_\nu n(\lambda)\right\rangle$,
defined in the space of control parameters $\lambda$, the counter-diabatic Hamiltonian is characterized by $\dot{\lambda}$.
Therefore, it is straightforward to expect a connection between the STA excess of work fluctuations and a quantum geometric quantity.
As shown in Sec.~\ref{sec2b}, this connection is described by
the equality in Eq.~(\ref{eq_15})~\cite{FunoPRL17}, which is further simplified to be Eq.~(\ref{eq_17}) for a single-qubit system.

\begin{figure}[t]
\centering
\includegraphics[width=0.65\columnwidth]{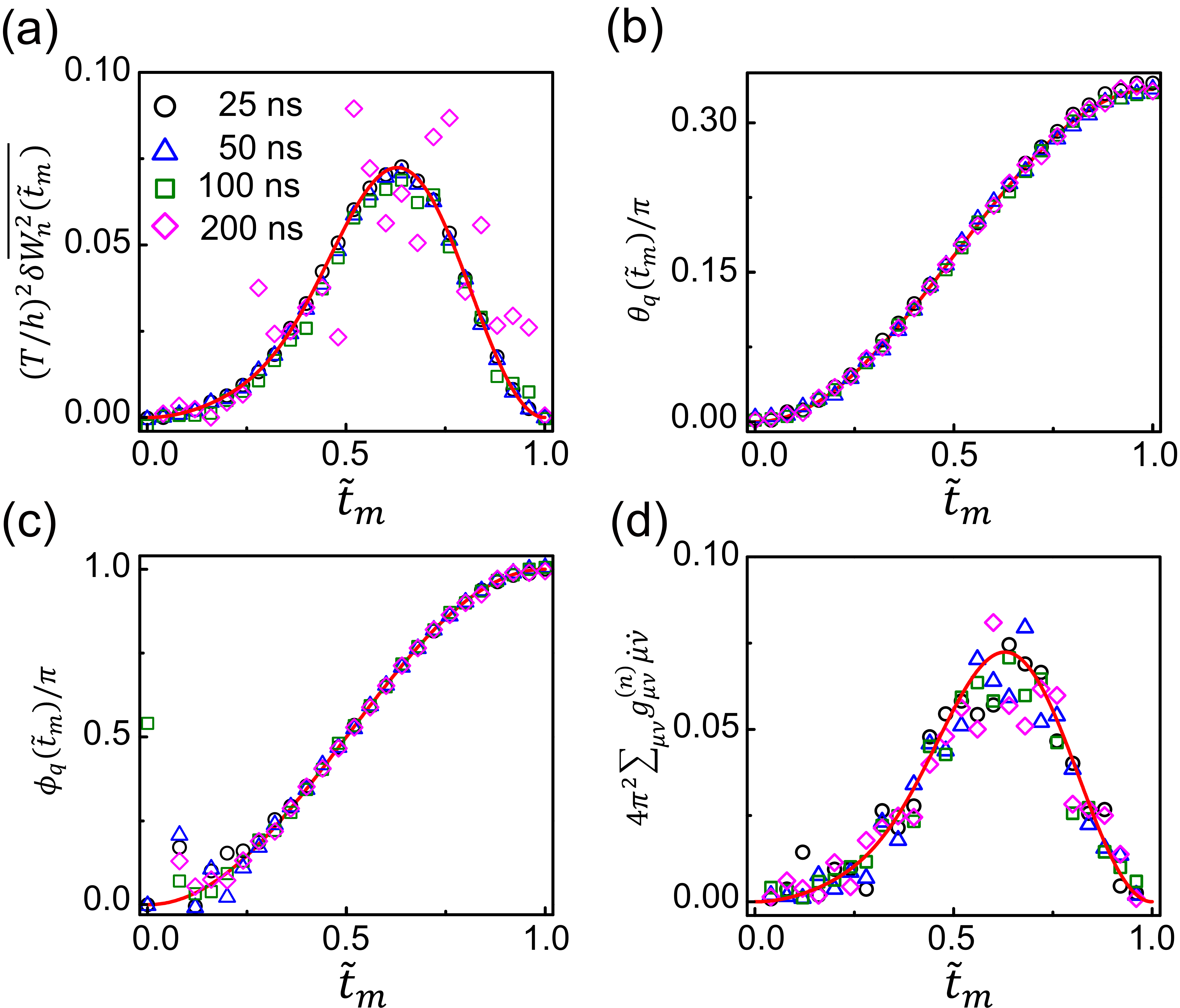}
\caption{Experimental verification of Eq.~(\ref{eq_17}) with $|n(0)\rangle=|s_\uparrow(0)\rangle$.
 (a) The rescaled STA excess of work fluctuations, $T^2 \overline{\delta W_n^{(2)}(\tilde{t}_m)}$.
 (b-d) The geometric tensor measurement:  (b) the polar angle  $\theta_q(\tilde{t}_m)$ and (c) the azithumal angle $\phi(\tilde{t}_m)$ of the qubit vector,
 and (f) the result of $4\pi^2 [dl(\tilde{t}_m)/d\tilde{t}_m]^2$, where the expression of $[dl(\tilde{t}_m)/d\tilde{t}_m]^2$ is shown in Eq.~(\ref{eq_22}).
The experimentally determined results are shown in symbols while the theoretical predictions are shown in solid lines. The STA operation time
is set at $T=25$, 50, 100 and 200 ns.}
\label{fig_n05}
\end{figure}

To verify Eq.~(\ref{eq_17}), we first experimentally determine the STA excess of work fluctuations with the operation time $T$.
Since the second moments for both initial eigenstates are theoretically identical, only $\overline{W^2_\uparrow(\tilde{t}_m)}$
are inspected to reduce the dissipation-induced error. For each operation time ($T=25$, 50, 100 and 200 ns),
the adiabatic reference is taken from the result of $T=500$ ns, giving
\be
\delta\overline{W^2_\uparrow(\tilde{t}_m)}=\overline{W^2_\uparrow(\tilde{t}_m; T)}-\overline{W^2_\uparrow(\tilde{t}_m; T=500~\mathrm{ns})}.
\label{eq_20}
\ee
Equation~(\ref{eq_17}) shows that $\delta\overline{W^2_\uparrow(\tilde{t}_m)}$ is inversely proportional to the square
of the operation time. Accordingly, we plot the rescaled value, $T^2 \overline{\delta W_\uparrow^{(2)}(\tilde{t}_m)}$,
in Fig.~\ref{fig_n05}(a). Consistent with the theoretical prediction, all the data of $T^2 \overline{\delta W_\uparrow^{(2)}(\tilde{t}_m)}$
fall into a single curve. The amplified error with $T=200$ ns is due to a very small difference between
$\overline{W^2_\uparrow(\tilde{t}_m; T=200~\mathrm{ns})}$ and $\overline{W^2_\uparrow(\tilde{t}_m; T=500~\mathrm{ns})}$.

Next we experimentally probe the quantum geometric quantity on the RHS of Eq.~(\ref{eq_17}). For each operation time,
the STA evolution is interrupted frequently with an time interval of $\delta\tilde{t}_mT=0.02T$ and the qubit state
is measured in the fixed frame of $\{\left|\uparrow\right\rangle, \left|\downarrow\right\rangle\}$ by
the quantum state tomography (QST). The extracted parameters, $\{r_q(\tilde{t}_m), \theta_q(\tilde{t}_m), \phi_q(\tilde{t}_m)\}$
of the qubit vector describe the qubit evolution in response to the STA field. In an ideal scenario,
the qubit initialized at the up state $|s_\uparrow(0)\rangle$ follows the same evolution of $\bm B_0(t)/|\bm B_0(t)|$, giving
$\theta_q(\tilde{t}_m)=\theta(\tilde{t}_m)$ and $\phi_q(\tilde{t}_m)=\phi(\tilde{t}_m)$.
For each operation time, these two equalities are
confirmed in Figs.~\ref{fig_n05}(b) and~\ref{fig_n05}(c). Notice that a small measurement error in the off-diagonal element $\rho_{\uparrow\downarrow}(\tilde{t}_m)$
can lead to an amplified error in the phase determination if $\theta_q(\tilde{t}_m)$ is around zero.
The trajectory of the qubit vector  allows us to experimentally estimate the time derivatives,
\be
\frac{dx(\tilde{t}_m)}{d\tilde{t}} = \frac{x(\tilde{t}_m+\delta\tilde{t}_m)-x(\tilde{t}_m-\delta\tilde{t}_m)}{2\delta\tilde{t}_m}
\label{eq_21}
\ee
with $x(\tilde{t}_m)=\theta_q(\tilde{t}_m)$ and $\phi_q(\tilde{t}_m)$. The geometric quantity in Eq.~(\ref{eq_17}) is
calculated as
\be
\left[\frac{dl(\tilde{t}_m)}{d\tilde{t}_m}\right]^2 = \frac{1}{4}\left[\left(\frac{d\theta_q(\tilde{t}_m)}{d\tilde{t}}\right)^2+\sin^2\theta_q(\tilde{t}_m)\left(\frac{d\phi_q(\tilde{t}_m)}{d\tilde{t}}\right)^2\right].
\label{eq_22}
\ee
As shown in Fig.~\ref{fig_n05}(d), the data of $[dl(\tilde{t}_m)/d\tilde{t}_m]^2$ with different operation times also fall into a single curve
from the theoretical prediction, although experimental errors are found. Through the combination of the above two measurements,
we experimentally verify the general equality connecting the STA excess of the work fluctuations and the quantum geometric quantity
in a simple quantum system of a single qubit.

\section{Summary}
\label{sec5}

In this paper, we apply a superconducting Xmon qubit to experimentally verify
statistics of `shortcut to adiabaticity' (STA) work theoretically
proposed in Ref.~\cite{FunoPRL17}. The  counter-diabatic Hamiltonian $H_\cd(t)$
additional to the reference adiabatic Hamiltonian $H_0(t)$
induces a quantum uncertainty of work cost in a fast STA operation.
Following the frozen-Hamiltonian and frozen-population schemes, we experimentally
measure the instantaneous eigenenergies and the population of instantaneous
eigenstates in regard to the total STA Hamiltonian, $H(t)=H_0(t)+H_\cd(t)$. With an experimental
realization of the two-point measurement scheme, we obtain statistics
of STA work $W_n(t)$ for two initial conditions, $|n(0)\rangle=|s_\uparrow(0)\rangle$ and $|s_\downarrow(0)\rangle$,
respectively. A series of STA operation times, $T=25$, 50, 100, 200 and 500 ns,
are studied in our experiment. As the result of $T=500$ ns is used to approximate
the reference adiabaticity, we successfully verify the theory of STA work:
(1) the average work is invariant with the supplementation of the counter-diabatic Hamiltonian,
i.e.,  $\overline{W_{n}(t)}=\overline{W_{\ad; n}(t)}$; (2) the energy cost of $H_\cd(t)$
is reflected by an increased work fluctuation, i.e.,  $\overline{W^2_{n}(t)}\ge\overline{W^2_{\ad; n}(t)}$;
(3) the STA excess of work fluctuations, $\delta\overline{W^2_{n}(t)}=\overline{W^2_{n}(t)}-\overline{W^2_{\ad; n}(t)}$,
is connected with the quantum geometric tensor of the STA protocol, $\delta\overline{W^2_{n}(t)}
=\h^2\sum_{\mu\nu}g^{(n)}_{\mu\nu}\dot{\lambda}_{\mu}\dot{\lambda}_{\nu}$.
In the case of an evenly time compression, $H_0\!=\!H_0(\tilde{t}=t/T)$,
the relation of $\delta\overline{W^2_{n}(t)}\propto T^{-2}$ is verified for the STA excess of work fluctuations.
This paper demonstrates the experimental availability of exploring quantum thermodynamics with a high-quality
superconducting Xmon qubit device.

\begin{acknowledgements}

We would like to thank Adolfo del Campo for helpful discussions.
The work reported here is supported by the National Basic Research Program of China (2014CB921203, 2015CB921004),
the National Key Research and Development Program of China (2016YFA0301700, 2017YFA0304303),
the National Natural Science Foundation of China (NSFC-11374260, 21573195, 11625419, 11474177),
the Fundamental Research Funds for the Central Universities in China, and the Anhui Initiative in Quantum Information Technologies (AHY080000).
This work was partially carried out at the University of Science and Technology of China Center for Micro and Nanoscale Research and Fabrication.

\end{acknowledgements}

\appendix
\section{Theoretical Proof of the STA Work Equalities}
\label{app1}

In this Appendix, we demonstrate the theoretical derivation of the two STA equalities in Eqs.~(\ref{eq_13}) and~(\ref{eq_14}). Our
derivation is slightly different from its original version in Ref.~\cite{FunoPRL17}.

With the reference Hamiltonian, $H_0(t)=\sum_m \varepsilon_m(t)|m(t)\rangle\langle m(t)|$, and the counter-diabatic Hamiltonian,
$H_\cd(t)=i\hbar \sum_m \calP^\perp_m(t) |\partial_t m(t)\rangle\langle m(t)|$, the $k$-th instantaneous eigenstate $|\psi_k(t)\rangle$ of the total Hamiltonian $H(t)=H_0(t)+H_\cd(t)$
satisfies
\be
\sum_m \langle m(t)|\psi_k(t)\rangle \left[\varepsilon_m(t)|m(t)\rangle+i\hbar \calP^\perp_m(t) |\partial_t m(t)\rangle\right]  = E_k(t) |\psi_k(t)\rangle,
\label{eq_app1}
\ee
and
\be
\left[E_k(t)-\varepsilon_n(t)\right]\langle n(t)|\psi_k(t) = i\hbar \sum_m \langle n(t)|\calP^\perp_m(t) |\partial_t m(t)\rangle.
\label{eq_app1a}
\ee
For the first moment of the STA work, its deviation from the adiabatic result is given by
\be
\overline{W_n(t)}-\overline{W_{\ad; n}(t)} &=& \sum_k \left[E_k(t)-\varepsilon_n(t)\right]P_{k|n}(t) \no \\
&=&\sum_k \left[E_k(t)-\varepsilon_n(t)\right] \langle n(t)|\psi_k(t)\rangle \langle \psi_k(t)|n(t)\rangle.
\label{eq_app2}
\ee
By substituting Eq.~(\ref{eq_app1a}) into Eq.~(\ref{eq_app2}), we obtain
\be
\overline{W_n(t)}-\overline{W_{\ad; n}(t)} &=&i\hbar\sum_k \sum_m \langle n(t)|\calP^\perp_m(t)|\partial_t m(t)\rangle\langle m(t)|\psi_k(t)\rangle\langle \psi_k(t)|n(t)\rangle \no \\
&=& i\hbar\langle n(t)|\calP^\perp_n(t)|\partial_t n(t)\rangle,
\label{eq_app3}
\ee
where the unitary operator, $\calI=\sum_k|\psi_k(t)\rangle\langle\psi_k(t)|$, and the orthogonality, $\langle m(t)|n(t)\rangle=\delta_{m, n}$, are used.
With the orthogonal projection, $\langle n(t)|\calP^\perp_n(t)=0$, the conservation of the average STA work in Eq.~(\ref{eq_13a}) is derived.
For the second moment of the STA work, its deviation from the adiabatic result is given by
\be
\overline{W^2_n(t)}-\overline{W^2_{\ad; n}(t)} &=& \sum_k \left[E_k(t)-\varepsilon_n(t)\right]^2P_{k|n}(t) \no \\
&=&\sum_k \left[E_k(t)-\varepsilon_n(t)\right]^2 \langle n(t)|\psi_k(t)\rangle \langle \psi_k(t)|n(t)\rangle.
\label{eq_app5}
\ee
Similarly, we substitute Eq.~(\ref{eq_app1a}) into Eq.~(\ref{eq_app5}), which leads to
\be
\overline{W^2_n(t)}-\overline{W^2_{\ad; n}(t)}&=& \hbar^2\sum_{m}\langle \partial_t m(t) |\calP^\perp_{m}(t)\calP_n(t)\calP^\perp_{m}(t)|\partial_t m(t)\rangle \no \\
&=&\hbar^2\sum_{m}\langle \partial_t m(t) |\calP_n(t)|\partial_t m(t)\rangle-\hbar^2\langle \partial_t n(t) |\calP_n(t)|\partial_t n(t)\rangle,
\label{eq_app6}
\ee
with the usage of $\calP_n(t)=|n(t)\rangle \langle n(t)|$ and $\calP^\perp_{m}(t)\calP_n(t)\calP^\perp_{m}(t)=\calP_n(t)-\delta_{m, n}\calP_n(t)$.
Next we take the time derivative over the unitary operator, $\calI=\sum_{m^\pr} |m^\pr(t)\rangle\langle m^\pr(t)|$, and obtain the follow equality,
$\sum_{m^\pr}|\partial_t m^\pr(t)\rangle\langle m^\pr(t)|=-\sum_{m^\pr}|m^\pr(t)\rangle\langle \partial_t m^\pr(t)|$. An inner product with the $m$-th reference eigenstate $|m(t)\rangle$
gives rise to
\be
|\partial_t m(t)\rangle &=& -\sum_{m^\pr}|m^\pr(t)\rangle\langle \partial_t m^\pr(t)|m(t)\rangle.
\label{eq_app8}
\ee
With the projection, $\calP_n(t)|m^\pr(t)\rangle=\delta_{m^\pr, n}|n(t)\rangle$, Eq.~(\ref{eq_app6}) is reorganized into
\be
\overline{W^2_n(t)}-\overline{W^2_{\ad; n}(t)}&=& \hbar^2\sum_{m} \langle \partial_t n(t)| m(t)\rangle \langle m(t) |\partial_t n(t)\rangle-\hbar^2\langle \partial_t n(t) |\calP_n(t)|\partial_t n(t)\rangle \no \\
&=& \hbar^2 \langle \partial_t n(t) |\calP^\perp_n(t)|\partial_t n(t)\rangle.
\label{eq_app9}
\ee

\end{document}